\documentstyle[12pt]{article}
\begin{document}

\begin{titlepage}
\begin{center}
July 1993
\hfill LBL-34561 \\
 \hfill UCB-PTH-93/23 \\
\vskip 0.2 in
{\large \bf THE WEAK SCALE MEASUREMENTS CONSTRAINTS ON STRING MODELS
 }\footnote{This work was supported by the
Director, Office of Energy Research, Office of High Energy and Nuclear
Physics, Division of High Energy Physics of the U.S. Department of Energy
under contract DE-AC03-76SF00098 and in part by the National Science
Foundation under grant PHY--90--21139.}\\
\vskip .2 in
{
{\bf Rulin Xiu} \\
\vskip 0.03 cm

{\it Department of Physics, University of California\\ and\\
Physics Division, Lawrence Berkeley Laboratory, 1 Cyclotron Road\\
Berkeley, CA 94720 \\ } }
\end{center} \vskip 0.05 in \begin{abstract}
The result of the modular invariant one-loop string effective coupling constant
for a large class of models
 is used to discuss the weak scale measurement constraints on  superstring
models.
Superstring models with intermediate gauge symmetry breaking  are proposed to
account for the grand gauge unification
scale $M_{GUT} = 10^{16} GeV$, which is  deduced from the precision weak scale
measurement in the minimal SUSY
 unification model.
  Gaugino condensation induced Hosotani mechanism  is suggested for the
intermediate
gauge symmetry breaking. In this scheme,  supergravity,
gauge symmetry and spacetime compactification are induced by one
nonperturbative dynamics: gaugino condensation.
The string scale $\mu_S$ is calculated in
 intermediate
SU(5) string models. A
relation between the intermediate gauge symmetry breaking scale $M_I\, = \,
M_{GUT}$
and string scale is found to be $ \mu_s^3 = M_Im_P^2 $. It is argued that this
relation is a general result for
all the superstring models with the intermediate gauge symmetry breaking.
The level one  minimal SUSY left-right string models are shown to be excluded
by weak scale measurement. The
constraints on the ratio of affine levels is worked out.
  It is demonstrated that string unification is more restricted by the
experiments than any other unification schemes are.
\end{abstract}
\end{titlepage}

\pagestyle{empty}
\newpage

\pagestyle{plain}
\pagenumbering{arabic}

\catcode`\@=11

\def\thesection{\arabic{section}.}
\def\thesubsection{\Alph{subsection}.}
\def\thesubsubsection{\Roman{subsubsection}.}

\def\ps@headings{\def\@oddfoot{}\def\@evenfoot{}}
\def\@oddhead{\hbox{}\hfill
\makebox[.5\textwidth]{\raggedright\ignorespaces --\thepage{}--
        \hfill }}
\def\@oddhead{\hbox{}\hfill --\thepage{}-- \hfill}
\def\@evenhead{\@oddhead}

\ps@headings
\catcode`\@=12

\relax
\section{Introduction}

In this work, the constraints and implications from the weak scale measurements
on string models are examined.
In an explicitly worked out string model, where the
matter content is specified and
the vevs of the dilaton and the  moduli are determined \cite{kapno},
 the string threshold correction can be calculated and
 the weak scale coupling constants can  be derived.
 But at the present stage of string phenomenology, we are far from
making these kinds of predictions. We are more interested in ``taming the zoo
of
string models'' with the help of low energy phenomenology.

 We use the modular invariant one loop string effective gauge coupling constant
derived in \cite{mary} and the analyzing method developed in
\cite{lin}. The computation in \cite{mary} indicates that the string scale
$\mu_s$ is the scale of two-loop gauge coupling
unification, up to the possible additional, modular-dependent threshold
corrections.
 Although this result  is computed explicitly for $E_8$ orbifold models,
I argue that, except for the explicit form of the string threshold corrections,
 it  covers a
large class of models including orbifold models, free fermion models and
Calabi-Yau models with arbitrary
  gauge interaction
content, since the derivation  is quite general which does not depend on the
gauge interaction content in a model and the form of
K\"ahler potential used for calculation  is not just limited to orbifold
models.

It has been found \cite{lin} that the constraints  from the low
energy measurements on  string models are more restrictive than on any other
unification models.
So far  minimal superstring unification and especially the contribution from
the string threshold correction to the gauge coupling
unification have been discussed extensively \cite{mod,mix,unif,pap1,lin}.
In \cite{unif} the constraints on the modular weights of the quark, lepton and
Higgs superfields has been worked out. In
our previous paper \cite{lin}, we worked out the restriction on the particle
spectrum in a SUSY minimal string model without
string threshold
corrections. Our method of analysis developed in \cite{lin}  is different from
\cite{unif} in that we identify string scale
  $\mu_s = \alpha\tilde{m}_P$ with the unification scale $g^{-2}_a(\mu_s) =k_a
g^{-2}$ for a class of models for example $Z_3$,
 $Z_7$ orbifold models and our analysis is to second-order from the
 quantum field theory point of view. Our second-order analysis is
 not very effective for studying the constraints on models with string
threshold corrections $ \Delta_a$ since
in this case the ``unification'' relation: $g^{-2}_a({1\over2}g^2M_P^2) =k_a
g^{-2} - \Delta_a/4\pi$,
( $ M_P = 1/\sqrt{8\pi G_N}$ is called the reduced Planck scale )
is too complicated.
But to first order,
  one can recognize \cite{lin} $\tilde{m}_P = m_P/2e^{1/2}$, ( $ m_P =
1/\sqrt{G_N}$ is the Planck scale )
    as the first order unification
 scale. If one assumes that
string threshold corrections $\Delta_a$
 are larger than two loop corrections, our analysis arrives at the relation:
\begin{eqnarray}\alpha_{a(\mu)}^{-1} = k_a\alpha^{-1} +
\frac{b_a}{2\pi}\,\ln\,\frac{\tilde{m}_P}{\mu} + \Delta_a
\end{eqnarray}
These put much stronger constraints on a string model.

In the following,  I will first review previous discussions on the minimal SUSY
string models and then  I try to make a complete
discussion on the different string models
 that  can give rise to $M_{GUT} = 10^{16} GeV$ ( the result deduced from the
precision weak scale measurement in the minimal SUSY
 unification model ).
Superstring models with intermediate gauge symmetry breaking  are proposed in
which SU(5), SO(10) or $E_6$
 gauge symmetry is broken at an
intermediate scale \( M_I = M_{GUT} = 10^{16}GeV \) .
The string scale $\mu_s$ is calculated from $M_{GUT} = 10^{16}GeV$ and $
\alpha_{(M_{GUT})} \simeq 25$ for the minimal  SU(5)
superstring models
  and  a relation between string scale, intermediate gauge symmetry breaking
scale and Planck scale is found to be:
$ \mu_s^3 = M_Im_P^2 $. It is argued that this relation is a general result for
superstring models with
 intermediate gauge symmetry breaking regardless of the dynamical gauge
symmetry breaking scheme or the gauge unification group.
 We also suggest a possible dynamic gauge symmetry breaking scheme for this
kind of models:
gaugino condensation induced Hosotani
Mechanism.
In this scheme, supergravity breaking, spacetime compactification and gauge
symmetry breaking may be triggered by one
nonperturbative effect: gaugino condensation.  It is also found that there is
some gaugino condensation induced
dynamical SUSY breaking model may give rise to $M_{GUT} = 10^{16}GeV$ and even
$M_{SUSY} = 1 TeV$.
 Then the minimal SUSY left-right string models (MSLRSM)
and the minimal left-right SUSY string models (MLRSSM) are discussed.
By the minimal left-right model, we mean the model restores left-right symmetry
at some
intermediate scale $M_R$ and it has the standard fermion content plus right
handed neutrino and
two types of Higgs similar to the model in \cite{lr}. In the MSLRSM, the SUSY
breaking scale $M_{SUSY}$ is less than
$M_R$. In the MLRSSM, $ M_{SUSY}$ is greater than $M_R$.
We find that just like in the level one minimal SUSY
string unification models discussed in \cite{lin}, in the level one  MSLRSM
 with an arbitrary number of Higgs
the weak scale measurements require  extra heavy  triplet fermions in the
model. But for level one MLRSSM there is
 no such requirement. This result indicates that the level one left-right model
constructed in \cite{font} should be a left-right
 SUSY string model.

\section{The Implications Of $M_{GUT} = 10^{16} GeV$ For Superstring Models}

   The precision weak scale measurement seems to indicate that the minimal
SUSY grand unification model leads to a good agreement with a single
unification scale of
\( M_{GUT} = 10^{16\pm 0.3}GeV \) \cite{unifi}  and a best fit for \( M_{SUSY}
\)
 around 1 Tev. It is amazing that this SUSY Grand unification idea works
perfectly and predicts correctly
the present experimental value for $\sin^2\theta_w(m_z)$, once one sets
$M_{SUSY}$ around the TeV
 scale and use the actual values for $\alpha_s(M_Z)$ and $\alpha_{em}(M_Z)$.
The bottom to tau mass ratio is also predicted
correctly; the proton lifetime is predicted to be above the present
experimental limits.
    This analysis gives
\( \alpha_{(M_{GUT})}^{-1} \simeq 25 \) .
This does not agree with the string unification condition $g^{-2}_a({1\over
2}g^2M_P^2) =k_a g^{-2} - \Delta_a/4\pi$ unless one  tunes the gauge coupling
constants and string threshold correction to   $10^{-6}$
if we identify \( \mu_{s} \) with \( M_{GUT} \).
  This straightforward analysis
 has been considered to
contradict with string phenomenology.
Many kinds of string models
have been explored to show that by putting in  heavy fermions and
string threshold
corrections or by considering higher affine level string models,
  string
models can confront  weak scale measurements\cite{lin,mod,far,flip}.
But in these string
models, one spoils the beautiful experimental indication of gauge interaction
 unification at $M_{GUT} = 10^{16} GeV$ . There are a few kinds of string
models one can come up with to
account for this experimental result from the string phenomenology point of
view.
For example, one can assume  for some string models  that at the string scale
one has different gauge couplings:
$g^{-2}_a({1\over
2}g^2M_P^2) =k_a g^{-2} - \Delta_a/4\pi$,  and when they evolve down to
$10^{16}GeV$ the coupling constants become
equal accidently. For this kind of models,
one gets three constraints:
\begin{eqnarray} k_a\alpha^{-1} +
\frac{b_a}{2\pi}\,\ln\,\frac{\tilde{m}_P}{M_{GUT}} + \Delta_a + \delta_a^{(2)}
 = \alpha_{a(M_{GUT})}
\simeq \,25
\end{eqnarray}
 One can also have some string models that have large gauge symmetry group, for
example SU(5), SO(10) or $ E_6$
 at string scale, and it is broken by some kind of
dynamics at $M_{GUT} = 10^{16} GeV$.
 So  in this kind of models, \( M_{GUT} \) is  the intermediate  gauge
symmetry
  breaking scale instead of the grand gravity and gauge interaction unification
  scale, {\it i.e.}  string scale.
    In the Higgs mechanism of gauge symmetry breaking,  for the string
  SU(5) model to have an intermediate gauge symmetry breaking scale, it has to
  be of a higher affine algebra level, since the 24 representation of scalars
which is
  necessary to
  break SU(5) is not allowed in the level one SU(5) string model, or one has to
appeal to other
  gauge symmetry breaking mechanism as will be discussed later.
  The building of
  higher affine level string models has been discussed in \cite{high}, but no
such model has been
worked out explicitly.  Here we discuss the simplest minimal
  supersymmetric string SU(5) models which have only three generations of
representation 5 and
  representation 10 fermions and one representation 24 higgs ($\Phi$) and one
representation 5 higgs (H)
  above the intermediate scale
  \( 10^{16} GeV \). In this case,
\begin{eqnarray}
b_5 & = & -15 + 2n_g + 5n_H + \frac{1}{2}n_{\Phi}  =  -3.5,  \end{eqnarray}
Here $n_g, n_H$ and $n_{\Phi}$ is the number of generations of the fermions,
the number of H higgs and the number of
$\Phi$ higgs respectively.
 \begin{eqnarray}  b_{55} & = & 46.4\,n_g + 150\,n_H + 9.8\,n_{\Phi} - 150  =
149.
\end{eqnarray}
Using the running coupling constant formula derived in \cite{lin}, one gets the
relation:
\begin{eqnarray}\alpha^{-1}_5 + \frac{1}{4\pi}(b_5 + b_{55}b^{-1}_5)\,
\ln\,\alpha_5 = \alpha^{-1}_{(M_{GUT})} - \frac{b_5}{2\pi}\,\ln\,\frac{\tilde
{m}_P}{M_{GUT}} + \frac{C^G_5}{2\pi}\,\ln2 + \nonumber \\
\frac{1}{4\pi}b_{55}b^{-1}_5\,\ln\,\alpha_{(M_{GUT})}
- \frac{b_5}{4\pi}\,\ln\,k - \Delta
\end{eqnarray} where $k$ is the affine level and $\alpha_5$ is the SU(5) gauge
coupling constant at the string scale.
Putting $\alpha^{-1}_{(M_{GUT})} = 25$ and $M_{GUT} = 10^{16}GeV$, one gets:
\begin{eqnarray} \alpha^{-1}_5 - 3.67\,\ln\,\alpha_5 & = & 39.74 + 0.28\,\ln\,k
- \Delta
\end{eqnarray}
In the case k=2 and with no string threshold correction {\it i.e.} $\Delta =
0$,
one gets: \( \alpha_5^{-1} = 27.74 \)  and \( \mu_s = 0.99\times10^{18} GeV\).
So under the assumption that the intermediate SU(5) gauge symmetry breaking
scale is
  the unification scale deduced from the experiment in
the minimal SUSY SU(5) model, the string scale can be calculated and is around:
$\mu_s = 10^{18} GeV $. For the level 3 model, $ k=3 $, the string scale is
calculated to be $ \mu_s = 1.21\times 10^{18}GeV$, for
$ k = 4$, $ \mu_s = 1.40\times 10^{18}GeV$.
 An interesting relation observed in these models is : $ \mu_s^3 =  M_{I}m_P^2
$.
This identity holds approximately for the intermediate SU(5) models with
different affine levels.
For the string models with the string threshold correction, the relation still
roughly holds.
Here we only discuss a very special case. Suppose the string threshold
correction cancels out all the
renormalization effects on the coupling constant from $M_{GUT}$ to $\mu_s$,
that is
 $ \alpha_{(M_{GUT})} = \alpha_5 = k^{-1}\alpha $, then the string scale for
$k=2$ is:
\[ \mu_{s} = \tilde{m}_{P}\sqrt{\alpha} = 1.05\times 10^{18}GeV \]
 and for $ k = 3,\, \mu_s = 1.28\times 10^{18}GeV $.
{}From the equation (29), one gets:
\begin{eqnarray}\Delta = 39.74 - \alpha^{-1}_5 + 3.67\, \ln \,\alpha_5 + 0.28\,
\ln\,k .
 \end{eqnarray}
For k = 2,  \( \Delta = 3.12 \).  Assuming that  $\Delta$ takes the form:
\begin{eqnarray}  \Delta \simeq \ln\,[|\eta(T)|^4( T + \bar{T} )]
\end{eqnarray}
then the modulus is of the order of \( T \simeq 1.77 \).

It has been shown that the left-right unification scheme \cite{lr} leads to the
unification scale $10^{15.20 \pm 0.25}GeV$.
In the same way
 intermediate left-right string models can be constructed.
The same discussion can be carried on this kind of models. It is not hard to
see the relation $\mu_s^3 = M_{I}m_P^2$ still roughly holds here. Because of
the relation
 $\mu_{s} = \tilde{m}_{P}\sqrt{\alpha}$,  $\mu_s$ is
always at the order of $10^{18}GeV$.  It is resonable to conclude that the
relation  $\mu_s^3 = M_{I}m_P^2$ holds approximately
for all the
superstring models with intermediate gauge symmetry breaking  regardless of the
gauge unification
 symmetry and symmetry breaking dynamics.
 This relation may give some hint at the dynamics of the symmetry breaking
scheme in
this kind of models. We will discuss about this possibility in the last part of
this section.

To avoid the unnaturalness of high affine level models and some
unattractiveness of the Higgs mechanism,
one can appeal to dynamical gauge symmetry breaking schemes. In the string
phenomenology, the nontrivial spacetime topology ( the
wilson line mechanism ) \cite{wilson} is introduced to obtain the realistic
gauge interactions. It was shown in \cite{host2} that
 in the multiply-connected  spacetime, given the boundary conditions, the
physically realized wilson lines and
 so the gauge symmetries should
 be dynamically determined. The problem of  dynamically determining
 gauge symmetries and gauge symmmetry breaking scales has hardly been discussed
in the string phenomenology so far. More
  work need to be done in this respect. Here we just propose a  simplistic
gauge symmetry breaking dynamics.
 In
the Hosotani mechanism discussed in \cite{host},  it was shown
that in a system of non-abelian gauge fields and fermions with minimal gauge
interaction, if part of the space-time is compact and
there exist  sufficiently heavy fermions, the gauge symmetry is dynamically
broken for some special cases.
If one argues that gaugino condensates may force spacetime to be compactified
\cite{dual,ferra}  and the onset of
spacetime compactification  breaks SU(5), SO(10) or $E_6$ grand unification
gauge interaction to the
standard model at the gaugino condensation scale by the  Hosotani mechanism,
one can relate  $M_{GUT} = 10^{16}
GeV$ to the  gaugino condensates scale which may  hopefully be determined from
stringy dynamics. It is interesting to note that in
this kind of  models  one can accomplish
gauge symmetry breaking, supergravity breaking and compactification of
spacetime
by one scheme.
 It also has the advantage of  relating  stringy dynamics directly  to the
``experimental result'' $M_{GUT} = 10^{16}
GeV$.

It is very encouraging to see that there has been  some stringy dynamical
model, for example
the string-inspired supergravity model at one loop constructed in \cite{d.m.}
that is capable of
 generating  gaugino condensation scale around $10^{16}GeV$. The model
contains three generations of matter in the untwisted sector, a consistent
parameterization of gaugino condensation effects,
and string-loop threshold effects needed to maintain modular invariance. It was
found that the scale degeneracy of the vacuum is
lifted at the one-loop level, allowing a determination of the fundamental
parameters of the effective low-energy theory. The
numerical results for $E_8$ ( in ref.\cite{d.m.}, it is stated as  $E_6$, which
is a misprint )
 hidden gauge symmetry gives gaugino
condensates scale: $\Lambda_R = 1.30\times10^{17} GeV$, and the coupling
constant at the string scale: $\alpha^{-1}  =  23.9$
which are very close to  the result we want. In the case of $E_6$ hidden gauge
symmetry, the numerical result will be closer.
  If one  follows the argument in \cite{softsusy},  soft
SUSY breaking scale at 1$TeV$ is very easy to obtain. It has been shown in
\cite{softsusy} that
 in some superstring models supersymmetry is broken in a hidden
sector but remains globally conserved in the observable sector of quarks and
gluons because of space-time duality.  The soft
supersymmetry is broken by anomalies and gaugino masses is generated through
two steps of radiative corrections. In this way,
the large hierarchy of scales is generated.  These show some  promise that
string dynamics might produce  $M_{GUT} = 10 ^{16}GeV$
and $M_{SUSY} = 1TeV$ for the superstring models with  intermediate gauge
symmetry breaking  proposed here.
 But to give a serious thought about this result,  a lot more work is needed.
We will pursue this line of ideas in the future.

\section{SUSY Left-Right Model and Left-Right SUSY Model}

In this part, I extend our discussion in \cite{lin} from minimal superstring
models to left-right superstring models.
The SUSY left-right models are another route of unification in which parity is
restored at some intermediate scale $M_R$. Apart from this appealing feature
of restoration of parity symmetry, the
 large mass differences between the left- and right-handed neutrinos
 can be accommodated in a natural way through the see-saw
mechanism.
A Z(3) orbifold SUSY left-right string model has been built explicitly in
\cite{font},
which is  a level one model without string threshold corrections.
In the
following we will  discuss the  minimal SUSY
 left-right string models (MSLRSM) and the minimal left-right SUSY string
models (MLRSSM)
 without  string threshold corrections.
By the minimal model we mean that the model has
the standard fermion content plus the right handed neutrino
 and two types of Higgs: \( \Delta_L = (0 1 0 2) \)
and \( H = (0 \frac{1}{2} \frac{1}{2} 0) \)
which is similar to the left-right model in \cite{lr}.
For the  MSLRSM we assume \( M_{SUSY} = M_T \equiv 1 TeV < M_R \equiv M_X \),
and for the MLRSST we suppose \( M_R = M_T \equiv 1 TeV < M_{SUSY}
 \equiv M_X \).
In either case, the running coupling constants can be written as:
\begin{eqnarray} \alpha^{-1}_{a(M_Z)} & = & k_{a}\alpha^{-1} +
\frac{b^{(I)}_{a}}{2\pi} \ln \frac{M_{T}}{M_Z} +
\frac{b^{(II)}_{a}}{2\pi} \ln \frac{M_X}{M_T} +
 \frac{b^{(III)}_{a}}{2\pi} \ln \frac{\tilde{m_p}}{M_X}
+ \delta_{a}^{(2)} + \Delta_{a},  \end{eqnarray}
Our analysis given in Appendix computes the constraints on affine levels, it
shows  that
 for the minimal left-right SUSY model:
$\,\, \alpha^{-1} < 30.77, \,\, \frac{k_2}{k_3} < 1.31 $
and for minimal SUSY left-right model:
$\,\,  \alpha^{-1} \,<\, 21.98,\,\, \frac{k_2}{k_3} \,<\, 2.26$.
It also indicates  that the weak scale measurements exclude  the level one
MSLRSM.
 Extra heavy triplet
fermions have to be added to make this kind of models  work,
just as in the case of the level one minimal SUSY string unification model
discussed in \cite{lin}. But for MLRSSM no such requirement is forced.
All these results are done for arbitrary number of higgs.

\section{Conclusion}

 Different superstring
models that can account for $M_{GUT} = 10^{16} GeV$ are discussed.
Superstring models with intermediate gauge symmetry breaking
is proposed.  The string scale is calculated for minimal superstring models
with SU(5) intermediate gauge symmetry breaking.
It is found that the relation $\mu_s^3
= M_Im_P^2$   holds approximately for all kinds of  superstring  models with
intermediate gauge symmetry breaking.
In  string phenomenology,  symmetry breaking should happen dynamically and  all
the scales are related to each other
and
 can be  determined from one or two parameters. The experimentally deduced
relation we find will put strong constraints on the
possible  underlying symmetry breaking dynamics in these types of models.
It is also argued that gaugino condensation may serve as a scheme to  break
gauge symmetry through the Hosotani mechanism in this
kind of models and give $M_{GUT} = 10^{16} GeV$. It is  encouragingly noticed
that some
dynamical string models may generate $M_{GUT} = 10 ^{16}GeV$ and $M_{SUSY} =
1TeV$. Further detailed discussion will be given
  elsewhere.
 The constraints on the ratio of affine levels are worked out for minimal SUSY
left-right string model (MSLRSM) and minimal
 left-right SUSY string models (MLRSSM). For the level one model, we find that
the MSLRSM is excluded for an
 arbitrary number of higgs. This analysis indicates that the level one
left-right string model
 in \cite{font} should be the MLRSSM, {\it i.e.} the SUSY breaking scale should
be larger than the left-right symmetry
breaking scale.
All the analysis is carried out  using  one loop modular invariant string
effective  coupling constant computed in \cite{mary}.
 Because of the relation between the  string unification scale $\mu_s$ and  the
coupling constant at string scale,  we find
that string unification models are more restricted
than any other unification schemes by low energy measurements.

\appendix{Appendix}

The analysis starts with the running coupling constants:
\begin{eqnarray} \alpha^{-1}_{a(M_Z)} & = & k_{a}\alpha^{-1} +
\frac{b^{(I)}_{a}}{2\pi} \ln \frac{M_{T}}{M_Z} +
\frac{b^{(II)}_{a}}{2\pi} \ln \frac{M_X}{M_T} +
 \frac{b^{(III)}_{a}}{2\pi} \ln \frac{\tilde{m_p}}{M_X}
+ \delta_{a}^{(2)} + \Delta_{a},  \end{eqnarray}
Here
\begin{eqnarray}\delta_{a}^{(2)} & = &  \frac{b^{(III)}_a} {4 \pi}\,
\ln \alpha - \frac{C_a^G}{2\pi} \ln\,2 + \frac{1}{4\pi} \,
\sum_{c} b^{(I)}_{ac}b^{(I)-1}_c \ln\,\frac{\alpha_{c(M_T)}}{\alpha_{c(M_Z)}} +
\frac{1}{4\pi}
\sum_{c} b^{(II)}_{ac}b^{(II)-1}_c \ln\frac{\alpha_{c(M_X)}}{\alpha_{c(M_T)}}
+\nonumber \\
            &   &   \frac{1}{4\pi}
\sum_{c} b^{(III)}_{ac}b^{(III)-1}_c
\ln\frac{\alpha_{c(\mu_s)}}{\alpha_{c(M_X)}},
\end{eqnarray}
and \begin{equation}
\Delta_{a} = - 2\pi\sum_{I} \beta_{a}^{I} \ln[ \eta^{2}(it^{I})
\overline{\eta^{2}(it^{I})}(t+\bar{t})^{I}]. \end{equation}
Using the lowest order expression for \(\ln\,[\alpha_{a(M_T)}
/\alpha_{a(M_Z)}], \,\,\ln\,[\alpha_{a(M_X})/\alpha_{a(M_T)}] \)
 and \( \\ \) \( \ln\,[\alpha_{a(\mu_s)}/\alpha{_a(M_X)}]\), eq.(34) can be
written in the form :
\begin{equation}  k_{a}\alpha^{-1} + b_a x +
c_a \ln\alpha = B_a\end{equation}
Here
\begin{eqnarray}
b_a & \equiv & b^{(II)}_a - b^{(III)}_a +
\frac{1}{4\pi}\sum_{c}(b^{(II)-1}_cb^{(II)}_{ac}
- b^{(III)-1}_cb^{(III)}_{ac}) b^{(II)}_c\alpha_{c(M_{SUSY})} \nonumber \\
    & \simeq &  b^{(II)}_a - b^{(III)}_a ,\\
c_a & \equiv & \frac{1}{4\pi} ( b^{(III)}_a + \sum_{c} b^{(III)-1}_c
b^{(III)}_{ac} ), \\
B_a & \equiv & \alpha^{-1}_{a(M_Z)} - \frac{{b'^{(I)}}}{2\pi}\ln\frac{M_T}{M_Z}
      - \frac{b^{(III)}}{2\pi}\ln\frac{\tilde{M_P}}{M_T} +
\frac{1}{4\pi}
\sum_{c} b^{(III)}_{ac}b^{(III)-1}_c \ln\,\alpha_{c(M_T)} +
\frac{c^G_a}{2\pi}\ln2 \nonumber \\
     &   & \mbox{} + \frac{1}{4\pi}\sum_{c} b^{(III)}_{ac}b^{(III)-1}_c \ln\,
k_c \\
x & \equiv & \frac{1}{2\pi} \ln \,\frac{M_X}{M_T}
\end{eqnarray}
For the MSLRSM with $\Delta$ number of $\Delta$ Higgs and H number of H Higgs,
 we get three restrictions :
\begin{eqnarray}
\lefteqn{k_1\alpha^{-1} + ( 0.6-9\Delta ) x + \frac{1}{4\pi}( 6+9\Delta+
\frac{7+54\Delta}{6+9\Delta}+\frac{18+72\Delta}{H+2\Delta}-\frac{8}{3})\,
\ln\alpha }  \nonumber \\
&  =  & 57.25 \pm 0.11 - 5.71 \times (6 + 9\Delta)  \nonumber \\
&  & - (0.32\frac{7 + 54\Delta}{H + 2\Delta}
+ 0.27\frac{18 + 72\Delta}{H + 2\Delta} - 0.19 \times \frac{8}{3}),
\end{eqnarray}
\begin{eqnarray}
\lefteqn{k_2\alpha^{-1} + ( 1-H-2\Delta )\, x + \frac{1}{4\pi}( H+2\Delta+
\frac{3+12\Delta}{6+9\Delta}+\frac{18+10H+24\Delta}{H+2\Delta}-8)\,
\ln\,\alpha  } \nonumber \\
& = &  30.84 \pm 0.11-5.71 \times (H+2\Delta) + 0.22 \nonumber \\
& & - (0.32\frac{3+12\Delta}{6+9\Delta}
+0.27\frac{18+10H+24\Delta}{H+2\Delta}-0.19\times 8),
\end{eqnarray}
\begin{eqnarray}
\lefteqn{k_3\alpha^{-1}  + \frac{1}{4\pi}( -3 +
\frac{1}{6+9\Delta}+\frac{18}{H+2\Delta}-\frac{14}{3})\,
\ln\,\alpha  } \nonumber \\
& = &   11.23 \pm 0.79 + 5.71\times 3 + 0.33 \nonumber \\
& & - (0.32\frac{1}{6 + 9\Delta}
+ 0.27\frac{18}{H + 2\Delta} - 0.19\times \frac{14}{3})
\end{eqnarray}
and for the MLRSSM,
 the constraints are:
\begin{eqnarray}
\lefteqn{k_1\alpha^{(-1)} + ( 1-9\Delta )\, x +\frac{1}{4\pi}\,( 6 + 9\Delta +
\frac{7 + 54\Delta}{6 + 9\Delta} + \frac{18+72\Delta}{H+2\Delta}-\frac{8}{3})\,
\ln\alpha } \nonumber \\
& = & 57.25 \pm0.11-5.71\times (6+9\Delta) - \nonumber \\
& &(0.32\frac{7+54\Delta}{6+9\Delta}
+0.27\frac{18+72\Delta}{H+2\Delta}-0.19 \times \frac{8}{3})
\end{eqnarray}
\begin{eqnarray}
\lefteqn{k_2\alpha^{(-1)} + ( -7/3-H-2\Delta ) x +
\frac{1}{4\pi}( H+2\Delta+
\frac{3+12\Delta}{6+9\Delta}+\frac{18+10H+24\Delta}{H+2\Delta}-8)\,
\ln\alpha  } \nonumber \\
&  = &  30.84 \pm 0.11 - 5.71\times (H+2\Delta) + 0.22 \nonumber\\
    & & - (0.32\frac{3 + 12\Delta}{6 + 9\Delta}
+ 0.27\frac{18 + 10H + 24\Delta}{H + 2\Delta} - 0.19\times 8)
\end{eqnarray}
 \begin{eqnarray}
\lefteqn{k_3\alpha^{(-1)} -4 x + \frac{1}{4\pi}( -3 +
\frac{1}{6+9\Delta}+\frac{18}{H+2\Delta}-\frac{14}{3})\,
\ln\alpha } \nonumber \\
& = & 11.23 \pm 0.79 + 5.71\times 3 + 0.33 \nonumber \\
&   & - (0.32\frac{1}{6 + 9\Delta}
+ 0.27\frac{18}{H + 2\Delta} - 0.19\times\frac{14}{3})
\end{eqnarray}

First we work out the general constraints on the unification coupling constant
$\alpha$ and the affine levels.
For an arbitrary number of Higgs, we have
\begin{eqnarray*}
\frac{1}{2} & \leq & \frac{3+12\Delta}{6+9\Delta} \,\, \leq \,\, \frac{4}{3} \\
         10 & \leq & \frac{10H+24\Delta}{H+2\Delta} \,\, \leq \,\, 12
\end{eqnarray*}
and also,
\begin{eqnarray*}
0  & \leq &  x  \,\, \leq \,\, \frac{1}{4\pi}\,\ln\,\alpha + 5.71,
\end{eqnarray*}
Using these inequality, we get for the minimal left-right SUSY model:
\[ k_2\,\alpha^{-1} < 30.77,\,\,\,\,\,\, \alpha^{-1} < 30.77,\,\,\,\,\,\,
k_3\,\alpha^{-1}\,\, > \, 23.52, \,\,\,\,\,\, \frac{k_2}{k_3} < 1.31 \]
and for minimal SUSY left-right model:
\[ k_2\,\alpha^{-1}\,<\, 49.67,\,\,\,\,\,\,\, \alpha^{-1} \,<\, 49.67,
\,\,\,\,\,\,
   k_3\,\alpha^{-1}\,>\, 21.98,\,\,\,\,\,\,\, \frac{k_2}{k_3} \,<\, 2.26\]
Next we discuss the level one left-right models, i.e. models with \( k_2 = k_3
= 1 \). Taking eq. (43) - eq. (42), one gets
\begin{eqnarray} ( H + 2\Delta - 1 )\,x & = & - 2.37 \pm 0.90 + 5.71\times ( H
+ 2\Delta ) +\nonumber \\
                                         &   & \frac{1}{4\pi}( H+2\Delta+

\frac{2+12\Delta}{6+9\Delta}+\frac{10H+24\Delta}{H+2\Delta}-\frac{1}{3})\,\
                                                 \ln\,\alpha +  \nonumber \\
                                         &   & \mbox{}
(0.32\frac{2+12\Delta}{6+9\Delta}
                                                +
0.27\frac{10H+24\Delta}{H+2\Delta}-0.19\times \frac{10}{3})
\end{eqnarray}
Since there is at least one higgs, the left-hand side of the equation is equal
to or greater than zero. It is not hard to work out
that  the right-hand side is always less than zero. This means that within the
experimental error for an arbitrary
number of Higgs,  the MSLRSM would
not work. Extra SU(3) triplet fermions have to be added. In fact one can
calculate the constraints on
the  corrections to the running coupling constant
equation from the extra heavy fermions $\Pi_a$:
\begin{eqnarray*}  \Pi_3 - \Pi_2 & = & - ( H + 2\Delta - 1 )\,x  - 2.37 \pm
0.90 + 5.71\times ( H + 2\Delta ) +\nonumber \\
                                 &   &       \frac{1}{4\pi}( H+2\Delta+

\frac{2+12\Delta}{6+9\Delta}+\frac{10H+24\Delta}{H+2\Delta}-\frac{1}{3})\,
                                                 \ln\alpha +  \nonumber \\
                                 &   & \mbox{}
(0.32\frac{2+12\Delta}{6+9\Delta}
                                                +
0.27\frac{10H+24\Delta}{H+2\Delta}-0.19\times \frac{10}{3})\\
                                 & > & -2.37 - 0.9 + 5.71 +
\frac{1}{4\pi}(\frac{4}{3} + 12 + 1 - 8 + 3 +
                                        \frac{14}{3})\,\ln\,\alpha\\
                                &   & ( 0.32\,\times\frac{1}{3} + 0.27\,\times
10 - 0.19,\times \frac{10}{3} )\\
                                 & > & 4.61 + 1.11\,\ln\,\alpha\\
                                 & > & 4.61 - 1.11\, \ln\,30.77\end{eqnarray*}
so $ \Pi_3 - \Pi_2 > 0.81 $. For the MLRSSM, there is not such an inequality.
No extra fermions have to be added to satisfy
the weak scale measurement in this kind of models. Applying this analysis to
the level one left-right model in \cite{font}
one can conclude that in this model the left-right symmetry breaking scale is
smaller than the SUSY breaking scale.

 \vskip 28pt
\noindent{\bf Acknowledgement.}
\vskip 12pt
 I am grateful to Mary K. Gaillard for going through the paper many times and
making  all kinds of corrections.
I also want to thank her and
my fellow graduate student Steve Johnson for lots of illuminating discussions.

This work was supported in part  the
Director, Office of Energy Research, Office of High Energy and Nuclear Physics,
Division of High Energy Physics of the U.S. Department of Energy under Contract
DE-AC03-76SF00098 and in part by the National Science Foundation under grant
PHY-90-21139.

\end{document}